\documentclass[prl,aps,amssymb,twocolumn,superscriptaddress]{revtex4-2}
\usepackage{graphicx,xcolor,caption,subcaption,wrapfig}
\usepackage{amsmath,amsfonts,mathtools,bm,amsthm}
\usepackage{multirow}
\usepackage{lmodern}
\usepackage{tikz-cd}
\usepackage{braket}
\usepackage{hyperref}
\usepackage[makeroom]{cancel}

\newcommand{\beqn}{\begin{eqnarray}}
\newcommand{\eeqn}{\end{eqnarray}}

\begin{document}

\title{Density Matrix Geometry and Sum Rules}

\author{Guangyue Ji}
\affiliation{Department of Physics, Temple University, Philadelphia, Pennsylvania, 19122, USA}

\author{David E. Palomino}
\affiliation{Department of Physics, Temple University, Philadelphia, Pennsylvania, 19122, USA}

\author{Nathan Goldman}
\affiliation{Laboratoire Kastler Brossel, Coll\`ege de France, CNRS, ENS-Universit\'e PSL, Sorbonne Universit\'e, 11 Place Marcelin Berthelot, 75005 Paris, France}
\affiliation{International Solvay Institutes, 1050 Brussels, Belgium}
\affiliation{Center for Nonlinear Phenomena and Complex Systems, Universit\'e Libre de Bruxelles, CP 231, Campus Plaine, B-1050 Brussels, Belgium}

\author{Tomoki Ozawa}
\affiliation{Advanced Institute for Materials Research (WPI-AIMR), Tohoku University, Sendai 980-8577, Japan}

\author{Peter Riseborough}
\affiliation{Department of Physics, Temple University, Philadelphia, Pennsylvania, 19122, USA}

\author{Jie Wang}
\email{jiewang.phy@gmail.com}
\affiliation{Department of Physics, Temple University, Philadelphia, Pennsylvania, 19122, USA}

\author{Bruno Mera}
\email{bruno.mera@gmail.com}
\affiliation{Advanced Institute for Materials Research (WPI-AIMR), Tohoku University, Sendai 980-8577, Japan}
\affiliation{Instituto de Telecomunica\c{c}\~oes and Departmento de Matem\'{a}tica, Instituto Superior T\'ecnico, Universidade de Lisboa, Avenida Rovisco Pais 1, 1049-001 Lisboa, Portugal}

\begin{abstract}
    Geometry plays a fundamental role in a wide range of physical responses, from anomalous transport coefficients to their related sum rules. Notable examples include the quantization of the Hall conductivity and the Souza-Wilkens-Martin (SWM) sum rule---both valid at zero temperature, independent of interactions and disorder. The finite-temperature generalization of the SWM sum rule has been explored in the literature, revealing deep connections to the geometry of density matrices. Building on recent advances in time-dependent geometric frameworks, we propose a time-dependent quantum geometric tensor for thermal density matrices. This formalism provides a unified interpretation of known sum rules within the framework of the fluctuation-dissipation theorem, further elucidating their fundamental geometric origin. In addition, it provides experimentally accessible methods to probe quantum geometry beyond the zero-temperature regime.
\end{abstract}

\maketitle
Geometry and topology play central roles in contemporary studies of quantum matter and quantum information. Among the key geometric quantities are the Berry curvature and the quantum metric, which govern a wide range of condensed matter phenomena—at zero temperature~\cite{Xiao_2010,KOLODRUBETZ20171}.

For interacting systems at zero temperature, when placed on a torus, the many-body ground state wavefunction, assuming it is non-degenerate, can be varied with respect to many-body momentum $\bm{K}$ [see Eq.~(\ref{eq:twistedBC})], giving rise to the quantum geometric tensor $\mathcal{Q}^{ab}(\bm{K})$. This Hermitian tensor decomposes into a real symmetric part—the many-body quantum metric—and an imaginary antisymmetric part—the many-body Berry curvature\footnote{When fractionalization phenomena occur, Eqn.~(\ref{defQ}) should be understood as the trace of the non-Abelian quantum geometric tensor defined over the degenerate ground state manifold.}:
\begin{equation}
    \mathcal{Q}^{ab}(\bm{K}) = G^{ab}(\bm{K}) + \frac{i}{2}\Omega^{ab}(\bm{K}).\label{defQ}
\end{equation}
The quantum metric $G^{ab}$ governs the localization length of the ground state, while the Berry curvature $\Omega^{ab}$ determines the static Hall conductivity~\cite{RestRMP,VanderbiltPolarization,MarzariVanderbilt,Resta_Localization,QianNiu99,peotta:torma:15,MohitPRX19}. Remarkably, both quantities can be probed through optical response functions, via a sum rule that connects the dissipative part of the optical conductivity to the quantum geometric tensor. Specifically, the Souza-Wilkens-Martin (SWM) sum rule captures the real part, while the celebrated Niu-Thouless-Wu result~\cite{Thouless85}—after applying the Kramers-Kronig relations—yields the imaginary part~\cite{Goldman24Pedagogical}:
\begin{equation}
    \int_0^{\infty} \frac{d\omega}{\omega} \, \sigma^{ab}_{D}(\omega, T=0) = \frac{\pi}{V}\frac{e^{2}}{\hbar} \mathcal{Q}^{ab}(\bm{K}).\label{SWM}
\end{equation}
Here, $\sigma_D^{ab}(\omega)$ denotes the dissipative (absorptive) component of the optical conductivity. In the thermodynamic limit, for gapped systems, the Berry curvature integrates to the many-body Chern number $\mathcal{C}/2\pi$.

Equation~\eqref{SWM} thus provides an experimentally viable route to probe the topology and geometry of quantum systems via dynamical response—especially in charge-neutral platforms such as cold-atom systems~\cite{Goldman_SciAdv17,GoldmanOzawa17PRB,GoldmanOzawa19PRR,GoldmanNP19,tan:etal:2019,klees:rastelli:cuevas:belzig:2020, Ozawa_2021, klees:cuevas:belzig:rastelli:2021,Goldman24Pedagogical,JianmingCai19,chen:li:palumbo:zhu:goldman:cappellaro:2022,WeiChenProbC22}.

At finite temperature, the thermal density matrix, rather than the ground state wavefunction, becomes the central object governing physical responses. The geometry of mixed states is a mathematically well-defined concept and has played an increasingly important role in quantum information science, particularly in areas such as quantum metrology~\cite{Uhlmann86,Uhlmann91}. The natural generalizations of the quantum metric and Berry curvature to mixed states are the quantum Fisher information matrix and the mean Uhlmann curvature, respectively. These quantities have been employed in a variety of contexts, including as entanglement witnesses~\cite{hyllus:etal:2012,toth:2012,Hauke2016,liu:yuan:lu:wang:2020} and as tools to probe thermal, quantum phase transitions and also dynamical phase transitions~\cite{zanaradi:giorda:cozzini:2007,zanardi:venuti:giorda:2007,paunkovic:vieira:08,viyuela:rivas:martin-delgado:2014,mera:vlachou:paunkovic:vieira:17,mera:vlachou:paunkovic:vieira:viyuela:18,mera:vlachou:paunkovic:vieira:17:boltzmann,amin:mera:vlachou:paunkovic:18,mera:19:review,amin:mera:paunkovic:vieira:19,silva:mera:paunkovic:21,mera:paunkovic:amin:vieira:2022}.

The zero-temperature sum rule in Eq.~\eqref{SWM} has been extended to thermal equilibrium systems at finite temperature, revealing deep connections between response functions and the geometry of the thermal density matrix~\cite{Hauke2016,Carollo_SciRep19}. The generalized sum rules take the form:
\begin{eqnarray}
    \int_0^{\infty} d\omega \, \tanh\left(\frac{\beta\omega}{2}\right) \frac{\Re[\sigma^{ab}_{D}(\omega)]}{\omega} &=&\frac{\pi}{V}\frac{e^{2}}{\hbar} \mathcal{F}^{ab}(\bm{K}),\label{gSWM1}\\
    \int_0^{\infty} d\omega \, \tanh^2\left(\frac{\beta\omega}{2}\right) \frac{\Im[\sigma^{ab}_{D}(\omega)]}{\omega} &=& \frac{\pi}{V}\frac{e^{2}}{\hbar}\mathcal{U}^{ab}(\bm{K}),\label{gSWM2}
\end{eqnarray}
where $\mathcal{F}^{ab}(\bm{K})$ and $\mathcal{U}^{ab}(\bm{K})$ denote, respectively, the quantum Fisher information matrix and the mean Uhlmann curvature associated with the thermal state at inverse temperature $\beta$; see below for definitions. Note that in the equations above, $\sigma^{ab}(\omega)$ also depends on $T$, but we omit this dependence from the notation except when explicitly evaluating at $T = 0$. We adopt this convention throughout the paper for other related quantities as well. 

The key insight behind the zero-temperature sum rule in Eq.~\eqref{SWM} is that the geometry of the ground state wavefunction can be expressed as a two-point correlation function of the center-of-mass (COM) position operator. Physically, the COM position operator captures the essence of optical transition processes~\cite{SWM00}. As a result, Eq.~\eqref{SWM} formally follows from the fluctuation-dissipation theorem, which relates two-point correlation functions (geometry) to dissipation (optical absorption).

Building on this perspective, a time-dependent quantum geometric tensor—defined as an unequal-time correlation function—was recently introduced~\cite{Raquel2403}. This framework successfully reproduces several geometric sum rules and elucidates their interrelations, particularly in the zero-temperature regime.

In contrast, the finite-temperature sum rules in Eqs.~\eqref{gSWM1} and~\eqref{gSWM2} were originally derived by comparing the spectral representations on both sides of the equations~\cite{Hauke2016,Carollo2018}. However, a unifying physical mechanism connecting the geometry of mixed states to thermal response functions has remained elusive.

In this work, we address this gap by introducing a time-dependent quantum geometric tensor for thermal density matrices. This allows us to systematically derive a family of geometric sum rules valid at arbitrary temperature in thermal equilibrium. Our formalism not only recovers the known results in Eqs.~\eqref{gSWM1} and~\eqref{gSWM2}, but also leads to new sum rules that offer a route to probe quantum information properties of the system via dynamical responses.

These results highlight the conceptual importance of viewing the geometry of both pure and mixed quantum states in terms of correlation functions.

\emph{Problem setting and response functions. --- } We consider a Hamiltonian $H(\bm\phi)$ smoothly parameterized by a real-valued, multi-dimensional parameter $\bm\phi = (\phi_1,\dots,\phi_d)$.
The eigenstates and eigenenergies of the grand-canonical Hamiltonian $K = H - \mu N$ are denoted by $|\Phi_m\rangle$ and $E_m$, respectively, where $m$ is a quantum number label and $N$ is the total particle number.

We assume that at each moment in time the system remains in thermal equilibrium. The corresponding thermal density matrix is given by
\begin{equation}
    \rho = \frac{e^{-\beta K}}{Z} = \sum_m p_m \ket{\Phi_m}\bra{\Phi_m}, \quad Z = \mathrm{Tr}(e^{-\beta K}),
\end{equation}
where the eigenvalues are $p_m \equiv \exp(-\beta E_m)/Z$, satisfying $\sum_m p_m = 1$. In the zero-temperature limit, the density matrix reduces to the ground state projector $\lim_{\beta \rightarrow \infty} \rho = |\Phi_0\rangle\langle\Phi_0|$~\footnote{The ground state may be degenerate, in which case the zero temperature limit is the projector onto the ground state subspace, appropriately normalized by its rank.}.

This work focuses on the geometry of thermal density matrices and their connection to response functions. We begin by reviewing the basic structure of linear response.

To leading order, we expand the Hamiltonian around $\bm\phi = 0$:
\begin{equation}
    H(\bm\phi) = H(0) + \sum_{i=1}^{d} \mathcal{O}^i \phi_i, \quad \mathcal{O}^i \equiv \left. \frac{\partial H}{\partial\phi_i} \right|_{\bm\phi = 0},
\end{equation}
where $\mathcal{O}^i$'s are the generalized current operators conjugate to $\phi_i$'s. Their time evolution in the Heisenberg picture is given by $\mathcal{O}^i(t) = e^{iKt/\hbar} \mathcal{O}^i e^{-iKt/\hbar}$.

According to Kubo's linear response theory, the susceptibility $\chi_{\mathcal{O}}^{jk}(t - t')$ relates the expectation value of the current at time $t$ to the perturbation applied at time $t'$:
\begin{equation}
    \langle\mathcal{O}^j(t)\rangle = \int_{-\infty}^{\infty} dt' \, \chi_{\mathcal{O}}^{jk}(t - t') \, \phi_k(t').
\label{eq: Kubo formula}
\end{equation}
This susceptibility is given by the retarded correlation function:
\begin{equation}
    \chi_{\mathcal{O}}^{jk}(t - t') = -\frac{i}{\hbar} \Theta(t - t') \langle [\mathcal{O}^j(t), \mathcal{O}^k(t')] \rangle,
\end{equation}
where $\Theta(t)$ is the Heaviside step function, and $\langle \cdot\rangle = \mathrm{Tr}(\rho \cdot)$ denotes thermal averaging.

The dissipative part of the response, which encodes absorption, is defined as
\begin{equation}
    \chi_{\mathcal{O}; D}^{jk}(t) \equiv \frac{1}{2i} \left[ \chi_{\mathcal{O}}^{jk}(t) - \chi_{\mathcal{O}}^{kj}(-t) \right].
\end{equation}

On the other hand, we consider the unequal-time two-point function of the current operators:
\begin{equation}
    \mathcal{S}_{\mathcal{O}}^{jk}(t) \equiv \langle \mathcal{O}^j(t) \mathcal{O}^k(0) \rangle.
\label{eq: two-point correlation function}
\end{equation}
In the frequency domain, this function is related to the dissipative response via the fluctuation-dissipation theorem:
\begin{equation}
    \hbar \, \chi_{\mathcal{O}; D}^{jk}(\omega) = -\frac{1}{2} \left(1 - e^{-\hbar\beta\omega} \right) \mathcal{S}_{\mathcal{O}}^{jk}(\omega).
\end{equation}

\emph{Geometry of density matrices. --- } We now turn to the geometry of thermal density matrices $\rho(\bm\phi)$ that depend smoothly on the multi-dimensional real parameter $\bm\phi$. The generalization of quantum geometry to mixed states considered here arises within Uhlmann’s geometric framework, which includes a principal bundle with connection---Uhlmann's connection---over the space of density matrices~\cite{Uhlmann86, Uhlmann91}.

Remarkably, the quantum geometric tensor can also be formulated without explicit reference to the underlying principal bundle structure. This alternative approach is based on the \emph{symmetric logarithmic differential} (SLD) of the density matrix. Given a smoothly varying density matrix $\rho(\bm\phi)$, the SLD is a one-form $\mathcal{L} = \mathcal{L}^j d\phi_j$ valued in Hermitian matrices and defined implicitly by the Lyapunov equation:
\begin{equation}
    d\rho = \mathcal{L} \rho + \rho \mathcal{L}.
\end{equation}
Using the spectral decomposition of $\rho$, one obtains the explicit expression:
\begin{align}
    \mathcal L^j =& \sum_m\frac{1}{2}\frac{\partial\log p_m}{\partial\phi_j} |\Phi_m\rangle\langle\Phi_m| \nonumber\\ 
    &+ i\sum_{m \neq n} \frac{p_n - p_m}{p_n + p_m} A^j_{mn} |\Phi_m\rangle \langle\Phi_n|,
\end{align}
where $A^j_{mn} \equiv -i\langle\Phi_m|\partial/\partial \phi_j|\Phi_n\rangle$ are the non-Abelian Berry connection coefficients. We refer to the first term as the classical part, $\mathcal L^j_C$, and the second term as the quantum part, $\mathcal L^j_Q$.

The finite-temperature quantum geometric tensor (FT-QGT), also known as the quantum Fisher tensor~\cite{Carollo2018}, is defined by
\begin{equation}
    \mathcal{S}^{jk} \equiv \mathcal{S}_{\mathcal L_Q}^{jk} = \langle \mathcal{L}^j_Q \mathcal{L}^k_Q \rangle = \mathcal{F}^{jk} + i\mathcal{U}^{jk},\label{defD}
\end{equation}
where $\mathcal{S}^{jk}$ is Hermitian. Its real part $\mathcal{F}^{jk} = \frac{1}{2} \langle\{\mathcal{L}^j_Q, \mathcal{L}^k_Q\}\rangle$ corresponds to the pullback of the Bures metric and is known as the quantum Fisher information (QFI) matrix (up to a factor of 4), while its imaginary part $\mathcal{U}^{jk} = \frac{-i}{2} \langle[\mathcal{L}^j_Q, \mathcal{L}^k_Q]\rangle$ is the mean Uhlmann curvature (MUC), as discussed in Refs.~\cite{Uhlmann86,Uhlmann91,Carollo2018}.

These quantities play central roles in quantum metrology, multipartite entanglement, and the study of phase transitions~\cite{TopDM15,Hauke2016,Carollo2018,KOLODRUBETZ20171,Fang25Fisher,ProbTopDM18,Xiaoguang16,XiangguangYangLeeZero23,XiaogangDiracMeasure23}. In the zero-temperature limit, assuming a finite gap, $\mathcal{S}^{jk}$ reduces to the standard quantum geometric tensor $\mathcal{Q}^{jk}$.

\emph{Time-dependent geometric tensor for density matrices and generating function for geometrical sum rules. --- } In this work, we introduce a time-dependent generalization of the quantum Fisher tensor, defined as the two-point correlation function of the quantum part of the SLD. Since geometry is central to our analysis, we simplify notation and denote it by $\mathcal{S}^{jk}(t)$:
\begin{equation}
    \mathcal{S}^{jk}(t) \equiv \mathcal{S}^{jk}_{\mathcal{L}_Q}(t) = \langle \mathcal{L}^j_Q(t)\, \mathcal{L}^k_Q(0) \rangle.
\end{equation}

In the appendix, we show that the two-point functions $\mathcal{S}_{\mathcal{O}}(t)$ and $\mathcal{S}_{\mathcal{L}_Q}(t)$ are closely related. This, together with the fluctuation-dissipation theorem, leads to a key result of this work:
\begin{equation}
    -\frac{1}{2\hbar} \mathcal{S}^{jk}(\omega) = \frac{\tanh^2\left( \frac{\hbar \beta \omega}{2} \right)}{1 - e^{-\hbar \beta \omega}} \ \frac{\chi^{jk}_{\mathcal{O}; D}(\omega)}{(\hbar \omega)^2},\label{generatingfunction}
\end{equation}
which relates geometry on the left-hand side to dissipation on the right-hand side. The relation can be seen as a generating function for sum rules for the time-derivatives of the time-dependent geometric tensor. Indeed, by multiplying both sides of Eq.~\eqref{generatingfunction} by positive powers of frequency $\omega^n$ and integrating over frequency, we obtain a family of generalized sum rules:
\begin{eqnarray}
    -\hbar \pi \, \mathcal{F}^{jk}_{(n)} &=& \int_0^{\infty} d\omega \, \tanh^{R_n}\left( \frac{\hbar \beta \omega}{2} \right) \frac{\Re[\chi^{jk}_{\mathcal{O}; D}(\omega)]}{\omega^{2 - n}}, \\
    -\hbar \pi \, \mathcal{U}^{jk}_{(n)} &=& \int_0^{\infty} d\omega \, \tanh^{I_n}\left( \frac{\hbar \beta \omega}{2} \right) \frac{\Im[\chi^{jk}_{\mathcal{O}; D}(\omega)]}{\omega^{2 - n}}.
\end{eqnarray}
Here, $f_{(n)}$ denotes the $n$-th time derivative of $f(t)$ evaluated at $t = 0$, i.e., $f_{(n)} = (i\partial_t)^n f(t)\big|_{t=0}$. The exponents $R_n$ and $I_n$ depend on the parity of $n$: if $n$ is even, then $R_n = 1$ and $I_n = 2$; if $n$ is odd, then $R_n = 2$ and $I_n = 1$.

The time-dependent quantum Fisher information matrix and mean Uhlmann curvature, which appear in these sum rules, are defined as the symmetric and antisymmetric parts of $\mathcal{S}^{jk}(t)$:
\begin{eqnarray}
    \mathcal{F}^{jk}(t) &=& \frac{1}{2} \left[ \mathcal{S}^{jk}(t) + \mathcal{S}^{kj}(t) \right], \\
    \mathcal{U}^{jk}(t) &=& \frac{1}{2} \left[ \mathcal{S}^{jk}(t) - \mathcal{S}^{kj}(t) \right].
\end{eqnarray}
It is important to note that since $\mathcal{S}^{jk}(t)$ is not Hermitian at finite $t$, both $\mathcal{F}^{jk}(t)$ and $\mathcal{U}^{jk}(t)$ are generally complex-valued.

\emph{Properties and exact bounds. --- } Crucially, any two-point function of the form $\mathcal{S}^{jk}_{\mathcal{O}}(t)$ in Eq.~\eqref{eq: two-point correlation function}, for a general operator $\mathcal{O}$, defines a positive semi-definite bilinear form. That is, for any set of test functions $c_j(t)$, the following inequality holds:
\begin{equation}
    \sum_{j,k} \int_{-\infty}^{\infty} dt \int_{-\infty}^{\infty} dt'\, c_j^*(t)\, \mathcal{S}^{jk}_{\mathcal{O}}(t - t')\, c_k(t')\geq 0. \label{positivity_Sjj}
\end{equation}
This inequality follows directly from the non-negativity of ${\rm Tr}[\rho\, X^\dagger X]$, where $X = \int_{-\infty}^{\infty} dt\, \sum_j c_j(t)\, \mathcal{O}^j(t)$. Since the time-dependent geometric tensor $\mathcal{S}^{jk}(t)\equiv \mathcal{S}_{\mathcal{L_{Q}}}(t)$ is itself a two-point function, it must satisfy the same positivity constraint, which imposes significant structural conditions. 

On the other hand, the dissipative part of the response function, $\chi^{jk}_{\mathcal{O}; D}(t)$, is also derived from two-point correlations and is known to be non-negative—physically reflecting the fact that dissipation rates are always non-negative. The generating function in Eq.~\eqref{generatingfunction} provides a natural bridge between these two aspects. In what follows, we focus on the positivity constraints imposed on the time-dependent geometric tensor, while the connection to dissipation is discussed in detail in the appendix.

The positivity of the time-dependent geometric tensor implies that its associated Hankel matrix $\mathcal{A}^{jk}_{mn}$ and all of its principal minors must be non-negative. The matrix elements of the Hankel matrix are given by
\begin{equation}
    \mathcal{A}^{jk}_{mn} = \mathcal{S}^{jk}_{(m+n)},
\end{equation}
where $\mathcal{S}^{jk}_{(m+n)} = (i\partial_t)^{m+n} \mathcal{S}^{jk}(t)\big|_{t=0}$ denotes the $(m+n)$-th time derivative at $t = 0$.

The positivity of the $k$-th principal minor implies that every $k \times k$ principal submatrix of $\mathcal{A}^{jk}_{mn}$ must be positive semi-definite. We now discuss the implications of this condition step by step.

First, the positivity of the order $k = 1$ principal minors implies that the diagonal entries of the Hankel matrix are non-negative. Since the time-dependent mean Uhlmann curvature vanishes under symmetrization, we conclude that for all $j$ and $m$, the Taylor coefficients of the time-dependent quantum Fisher information must be non-negative:
\begin{eqnarray}
\mathcal{S}^{jj}_{(2m)} = \mathcal{F}^{jj}_{(2m)} \geq 0.    
\end{eqnarray}

Next, fixing $m$, we consider the submatrix with entries labeled by indices $j$ and $k$. The positivity of this submatrix implies the determinant inequality
\begin{equation}
    \det \mathcal{F}_{(2m)} \geq \det \mathcal{U}_{(2m)},
\end{equation}
where the determinant is taken over the spatial indices $j$ and $k$. For $m = 0$, this reduces to the well-known determinant condition for the static Fisher tensor (see e.g., Ref.~\cite{Carollo2018}).

Furthermore, one can consider a $2 \times 2$ principal minor involving components at mixed orders, leading to a Cauchy-Schwarz-type inequality:
\begin{equation}
    \det\begin{bmatrix}
        \mathcal{S}^{jj}_{(2m)} & \mathcal{S}^{jk}_{(m+n)} \\
        \mathcal{S}^{kj}_{(m+n)} & \mathcal{S}^{kk}_{(2n)}
    \end{bmatrix} = 
    \mathcal{S}^{jj}_{(2m)}\, \mathcal{S}^{kk}_{(2n)} - \mathcal{S}^{jk}_{(m+n)}\, \mathcal{S}^{kj}_{(m+n)} \geq 0.
\end{equation}

We emphasize that all of these positivity constraints ultimately arise from the non-negativity of the dissipative response function~\cite{LFfundamentalbound24,LFquantumweight24,Raquel2403,Nagaosa_2025}, and reflect the fundamental constraint that geometry, as encoded in two-point functions of the SLD operators, must be physically consistent with dissipation and vice-versa.

\emph{Case study: optical conductivity sum rules. --- }
As our main case study, we consider optical conductivity. In this context, the relevant parameter space for defining geometric response is the space of twisted boundary conditions, as we now describe.

We consider a periodic system whose unit cell is spanned by basis vectors $\bm{a}_j$ for $j = 1, \dots, d$. The system as a whole is spanned by $N_j$ unit cells along the $j$-th direction, giving system-size vectors $\bm{L}_j = N_j \bm{a}_j$. The reciprocal lattice is spanned by vectors $\bm{b}^j$ satisfying $\bm{a}_i \cdot \bm{b}^j = 2\pi \delta_i^j$. We use Latin indices $i, j$ to label basis vectors, and $a, b$ to label Cartesian components.

The many-body wavefunction must satisfy twisted boundary conditions: when any particle is translated by $\bm{L}_j$, the wavefunction acquires a phase,
\begin{align}
    \Psi_m(\bm{r}_1,\dots, \bm{r}_n + \bm{L}_j,\dots, \bm{r}_{N_e}) 
    = e^{i \alpha_j} \Psi_m(\bm{r}_1,\dots, \bm{r}_{N_e}), \label{eq:twistedBC}
\end{align}
where $n = 1, \dots, N_e$ labels the particles, $\alpha_j$ denotes the twist angle, which is related to the many-body momentum $\bm K$ by $\alpha_j\equiv\bm K \cdot \bm L_j$. The vector $\bm{\alpha} = (\alpha_1, \dots, \alpha_d)$  consisting of twist angles defines our geometric parameter space~\footnote{These boundary conditions can be generalized in the presence of a magnetic field. In that case, the left-hand side of Eq.~\eqref{eq:twistedBC} involves a projective representation of the lattice $\Lambda$ generated by the vectors $\bm{L}_j$, realized via magnetic translation operators. The right-hand side remains unchanged.}. Since twist angles differing by $2\pi$ correspond to the same physical boundary condition, the twist-angle space is a $d$-dimensional torus.

The optical conductivity $\sigma^{ab}(t - t')$ characterizes the linear response of the current density $\langle \mathcal{J}^a(t) \rangle$ to a spatially uniform time-dependent electric field $E_b(t')$:
\begin{equation}
    \langle \mathcal{J}^a(t) \rangle = \int dt' \, \sigma^{ab}(t - t') \, E_b(t').
\end{equation}

The twisted boundary conditions can be absorbed into the electromagnetic vector potential via a gauge transformation. A time-dependent vector potential then introduces an electric field, establishing a connection with Eq.~\eqref{eq: Kubo formula} and the present formalism. See also Appendix~B3 for a more detailed discussion. 

The dissipative (absorptive) part of the optical conductivity is given by $\sigma^{ab}_D(\omega) = -\chi^{ab; D}_{\mathcal{J}}(\omega)/\omega$. From Eq.~\eqref{generatingfunction}, we then obtain the generating function for the optical conductivity:
\begin{equation}
    \frac{1}{2V}\frac{e^{2}}{\hbar}\mathcal{S}^{ab}(\omega)=\frac{\tanh^{2}(\hbar\beta\omega/2)}{1-e^{-\hbar\beta\omega}}\frac{\sigma_{D}^{ab}(\omega)}{\omega}\label{key}
\end{equation}
which, upon integrating over frequency, yields the zeroth-order geometric sum rules of Eqs.~\eqref{gSWM1} and~\eqref{gSWM2} as derived in Refs.~\cite{Carollo2018,Hauke2016}.

We now turn to a geometric sum rule for two-dimensional orbital magnetization at finite temperature. We define a time-dependent correlator
\begin{equation}
    \mathcal{M}(t) \equiv \varepsilon_{ab} \langle \mathcal{R}^a(t) \, \mathcal{J}^b(0) \rangle,
\end{equation}
which reduces to the orbital magnetization $\mathcal{M}$ in the static limit. Its Fourier transform satisfies
\begin{equation}
    \mathcal{M}(\omega) = -\frac{\hbar}{e} \varepsilon_{ab} \frac{2 \Im[\sigma^{ab}_D(\omega)]}{1 - e^{-\hbar \beta \omega}},
\end{equation}
and integration over frequency yields the sum rule:
\begin{equation}
    -\frac{\pi e}{\hbar} \mathcal{M} = \varepsilon_{ab} \int_0^\infty d\omega \, \coth\left( \frac{\hbar \beta \omega}{2} \right) \Im[\sigma^{ab}_D(\omega)],
\end{equation}
which generalizes the magnetic circular dichroic (MCD) sum rules derived by Kuneš and Oppeneer~\cite{Kunes_2000}, and by Souza, Vanderbilt, and Resta~\cite{Souza_Vanderbilt_2008, Resta_2020}, to finite temperature.

In the zero-temperature limit, the physical quantities appearing in the MCD sum rule become related:
\begin{equation}
    -\frac{\pi e}{\hbar} \mathcal{M}(T = 0) = \frac{\pi}{V} \frac{e^2}{\hbar} \varepsilon_{ab} \mathcal{U}^{ab}_{(1)}(T = 0) = \varepsilon_{ab} I^{ab},
\end{equation}
where $I^{ab} \equiv \int_0^\infty d\omega \, \Im[\sigma^{ab}_D(\omega)]$ is the optical integral introduced in Resta’s formulation~\cite{Resta_2020}. The final equality provides a geometric interpretation of $I^{ab}$ in terms of the mean Uhlmann curvature at $T=0$, i.e., the conventional Berry curvature. However, this relation breaks down at finite temperature. For instance, the sum rule for the finite-temperature mean Uhlmann curvature takes the form
\begin{equation}
    \frac{\pi}{V} \frac{e^2}{\hbar} \mathcal{U}^{ab}_{(1)} = \int_0^\infty d\omega \, \tanh\left( \frac{\hbar \beta \omega}{2} \right) \Im[\sigma^{ab}_D(\omega)],
\end{equation}
which is no longer proportional to either $\mathcal{M}$ or $I^{ab}$, highlighting the richer structure of geometric response at finite temperature. 

\emph{Conclusions. ---} 
In this work, we have introduced a time-dependent quantum geometric tensor for thermal density matrices, establishing a unified geometric interpretation of various sum rules within the framework of the fluctuation-dissipation theorem. The non-negativity of this tensor, rooted in its formulation as a correlation function, yields a systematic approach to deriving a range of exact bounds. Furthermore, our framework enables experimentally accessible methods to probe quantum geometric properties at finite temperatures, thereby extending the conceptual and practical reach of quantum geometry beyond the zero-temperature limit. 

\begin{acknowledgements}
\emph{Acknowledgement.---} We acknowledge Nishchhal Verma, Raquel Queiroz and Fang Xie for useful discussions.

B.~M. acknowledges support from the Security and Quantum Information Group (SQIG) in Instituto de Telecomunica\c{c}\~{o}es, Lisbon. This work is funded by FCT (Funda\c{c}\~{a}o para a Ci\^{e}ncia e a Tecnologia) through national funds FCT I.P. and, when eligible, by COMPETE 2020 FEDER funds, under Award UIDB/50008/2020 and the Scientific Employment Stimulus --- Individual Call (CEEC Individual) --- 2022.05522.CEECIND/CP1716/CT0001, with DOI 10.54499/ 2022.05522.CEECIND/CP1716/CT0001. N.~G. is supported by the ERC (LATIS project), the EOS project CHEQS, the FRS-FNRS Belgium and the Fondation ULB T.~O. acknowledges suport from JSPS KAKENHI Grant Number JP24K00548, JST PRESTO Grant No. JPMJPR2353.
\end{acknowledgements}

\bibliography{ref}

\clearpage
\onecolumngrid
\appendix

\tableofcontents

\section*{--- Appendix ---}
In this appendix, we supplement the main text with detailed derivations and comparisons. Section~A contrasts our results with those obtained by the Columbia group. Section~B summarizes the derivation of the geometric sum rules. Finally, Section~C elaborates on the inequalities associated with the time-dependent quantum geometric tensor.
\section{A: Comparison with Columbia Papers}
%
%
Refs.~\cite{Raquel2403, Raquel2406}, hereafter referred to as the Columbia papers, introduced a time-dependent quantum geometric tensor whose definition and spectral representation we briefly review below:
\begin{eqnarray}
    \text{Columbia:} \quad \mathcal{Q}^{ab}(t) &=& \mathrm{Tr}\left[\rho\, \mathcal{R}^a(t)\, (1 - \rho)\, \mathcal{R}^b(0)\right], \nonumber\\
    &=& \sum_{mn} p_m (1 - p_n)\, e^{i(E_m - E_n)t/\hbar}\, A^a_{mn} A^b_{nm},
\end{eqnarray}
where $\rho = \sum_m p_m |\Phi_m\rangle \langle \Phi_m|$ is the many-body thermal density matrix.

For comparison, we recall the definition of our time-dependent quantum geometric tensor and its spectral decomposition:
\begin{eqnarray}
    \text{This work:} \quad \mathcal{S}^{ab}(t) &=& \mathrm{Tr}\left[\rho\, \mathcal{L}^a_Q(t)\, \mathcal{L}^b_Q(0)\right], \nonumber\\
    &=& \sum_{mn} p_m \left( \frac{p_m - p_n}{p_m + p_n} \right)^2\, e^{i(E_m - E_n)t/\hbar}\, A^a_{mn} A^b_{nm}.
\end{eqnarray}
It follows immediately that $\mathcal{S}^{ab}(t) \neq \mathcal{Q}^{ab}(t)$ at any finite temperature $T \neq 0$. However, in the zero-temperature limit where $p_0 \rightarrow 1$ and $p_{m \neq 0} \rightarrow 0$, both expressions reduce to the same quantity.

A further distinction lies in the treatment of diagonal terms. The spectral representation of $\mathcal{S}^{ab}(t)$ includes only off-diagonal matrix elements due to the off-diagonal nature of $\mathcal{L}^a_Q$. In contrast, $\mathcal{Q}^{ab}(t)$ as defined by the Columbia group includes diagonal contributions of the form $p_m(1 - p_m)\, A^a_{mm} A^b_{mm}$.

As noted in Ref.~\cite{Raquel2403} and discussed in their Appendix, these diagonal terms introduce a subtle gauge dependence. Although the authors argue that, due to symmetry in exchanging the indices $a$ and $b$, such terms do not contribute to the antisymmetric part of $\mathcal{Q}^{ab}(t)$, they do contribute to its symmetric part. In contrast, our construction of $\mathcal{S}^{ab}(t)$ inherently avoids this issue by excluding diagonal terms altogether. The underlying reason is that our construction is based on the geometry of the space of density matrices—the base space in Uhlmann's principal bundle framework—which inherently preserves gauge invariance.
\section{B: Geometrical Sum Rules}
\subsection{B1: Linear response theory and fluctuation-dissipation theorem}
In this work, we consider a time-independent Hamiltonian $H[\bm\phi(t)]$ in thermal equilibrium, which is perturbed by a time-dependent, spatially uniform, real-valued perturbation field $\bm\phi(t)$ that couples to the total current operator $\mathcal{O}$ in the system. To leading order in perturbation theory,
\begin{equation}
    H(\bm\phi) = H(\bm\phi=0) + \delta H(t), \quad \delta H(t) = \sum_{j=1}^{d} \mathcal{O}^j \phi_j(t), \quad \mathcal{O}^j = \left.\frac{\partial H(\bm\phi)}{\partial\phi_j}\right|_{\bm\phi=0},
\end{equation}
where, in principle, the total current operator is a sum over local current operators throughout the system, $\mathcal{O}^j = \int_V d^d\bm{r}~\mathcal{O}^j_{\mathrm{loc}}(\bm{r})$. We consider finite periodic systems of total volume $V$ and adopt the following Fourier transform conventions: for a general function $f(\bm{r}, t)$,
\begin{equation}
    f(\bm{r}, t) = \frac{1}{V} \sum_{\bm{q}} \int_{-\infty}^{\infty} \frac{d\omega}{2\pi} e^{-i\omega t} e^{i\bm{q} \cdot \bm{r}} f(\bm{q}, \omega), \quad
    f(\bm{q}, \omega) = \int_V d^d\bm{r} \int_{-\infty}^{\infty} dt\, e^{i\omega t} e^{-i\bm{q} \cdot \bm{r}} f(\bm{r}, t),
\end{equation}
which follow from the identity resolutions $\sum_{\bm{L}} \delta(\bm{r} + \bm{L}) = V^{-1} \sum_{\bm{q}} e^{i\bm{q} \cdot \bm{r}}$ and $V \delta_{\bm{q}, \bm{0}} = \int_V d^d\bm{r}\, e^{-i\bm{q} \cdot \bm{r}}$. The thermodynamic limit corresponds to $V \to \infty$, in which $V^{-1} \sum_{\bm{q}} \to \int d^d\bm{q}/(2\pi)^d$.

Kubo's linear response theory dictates that the expectation value of the total current operator $\mathcal{O}^i$ in the perturbed state is, to leading order, given by the retarded response:
\begin{equation}
    J_{\mathcal{O}}^i(t) \equiv \mathrm{Tr}[\rho(t) \mathcal{O}^i] = \int_{-\infty}^{\infty} dt'\, \sum_{j=1}^{d} \chi^{ij}_{\mathcal{O}}(t - t') \phi_j(t'),
\end{equation}
where the susceptibility is defined as
\begin{equation}
    \chi^{jk}_{\mathcal{O}}(t - t') = -\frac{i}{\hbar} \Theta(t - t') \langle [\mathcal{O}^j(t), \mathcal{O}^k(t')] \rangle,
\end{equation}
with $\mathcal{O}^a(t) \equiv e^{itK/\hbar} \mathcal{O}^a e^{-itK/\hbar}$ the time-evolved operator in the Heisenberg picture, and $\langle \cdots \rangle \equiv \mathrm{Tr}(\rho \cdots)$ the thermal average with respect to the unperturbed equilibrium state $\rho = e^{-\beta K}/Z$.

For a spatially uniform, time-dependent, and real-valued perturbation $\bm{\phi}(t)$, the instantaneous dissipation rate is given by
\begin{equation}
    \frac{dW}{dt} = \frac{d}{dt} \mathrm{Tr}[\rho H(\bm\phi(t))] 
    = \mathrm{Tr}[\dot{\rho} H(\bm\phi)] + \mathrm{Tr}[\rho \dot{H}(\bm\phi)] 
    = \mathrm{Tr}[\rho \dot{H}(\bm\phi)] = \mathrm{Tr}[\rho\, \mathcal{O}^j \dot{\phi}_j(t)] 
    = J^j_{\mathcal{O}}(t) \dot{\phi}_j(t),
\end{equation}
where we have used $i\hbar \dot{\rho} = [H(\bm\phi), \rho]$. Integrating over time gives the total dissipation:
\begin{equation}
    \int_{-\infty}^{\infty} dt\, \frac{dW}{dt} = \int_{-\infty}^{\infty} dt \int_{-\infty}^{\infty} dt'\, \chi^{jk}_{\mathcal{O}}(t - t') \dot{\phi}_j(t) \phi_k(t') 
    = -\int_{-\infty}^{\infty}  \frac{d\omega}{2\pi }\,\omega\, \chi^{jk}_{\mathcal{O};D}(\omega) \phi_j^*(\omega) \phi_k(\omega),
    \label{app_general_dissipation}
\end{equation}
where we used the reality of $W(t)$ and $\phi_j(t)$ in the final equality. Here, the dissipative response is given by the anti-Hermitian part:
\begin{equation}
    \chi^{jk}_{\mathcal{O};D}(t) = \frac{1}{2i} \left[\chi^{jk}_{\mathcal{O}}(t) - \chi^{kj}_{\mathcal{O}}(-t)\right]. \label{app_dissipation_general}
\end{equation}

On the other hand, fluctuations are captured by the two-point correlation function,
\begin{equation}
    \mathcal{S}^{jk}_{\mathcal{O}}(t) \equiv \langle \mathcal{O}^j(t)\, \mathcal{O}^k(0) \rangle,
\end{equation}
whose spectral representation is
\begin{equation}
    \mathcal{S}^{jk}_{\mathcal{O}}(\omega) = 2\pi \hbar\sum_{mn} p_m\, \mathcal{O}^j_{mn} \mathcal{O}^k_{nm} \delta(\hbar\omega + E_m - E_n).
    \label{app:SpectrumRepS}
\end{equation}
By construction, $\mathcal{S}^{jk}_{\mathcal{O}}(\omega)$ is Hermitian at fixed $\omega$ and, because $\mathcal{S}^{jk}_{\mathcal{O}}(t)$ satisfies the Kubo–Martin–Schwinger (KMS) condition,
\begin{equation}
    \mathcal{S}^{jk}_{\mathcal{O}}(t) = \mathcal{S}^{kj}_{\mathcal{O}}(-t - i\hbar\beta),
\end{equation}
we have the following identities in the frequency domain:
\begin{equation}
    \mathcal{S}^{jk}_{\mathcal{O}}(\omega) = \left[\mathcal{S}^{kj}_{\mathcal{O}}(\omega)\right]^* = e^{\hbar\beta\omega} \mathcal{S}^{kj}_{\mathcal{O}}(-\omega).
\end{equation}

We define the symmetric and antisymmetric parts of the two-point function:
\begin{equation}
    \mathcal{S}^{jk}_{\mathcal{O};S/A}(\omega) \equiv \frac{1}{2} \left[\mathcal{S}^{jk}_{\mathcal{O}}(\omega) \pm \mathcal{S}^{kj}_{\mathcal{O}}(-\omega)\right] 
    = \frac{1}{2} \left(1 \pm e^{-\hbar\beta\omega}\right) \mathcal{S}^{jk}_{\mathcal{O}}(\omega),
\label{app_def_Ss_Sa}
\end{equation}
so that the KMS condition becomes the second identity above. The antisymmetric part is directly related to the dissipative response:
\begin{equation}
    \hbar\, \chi^{jk}_{\mathcal{O};D}(t) = -\mathcal{S}^{jk}_{\mathcal{O};A}(t),
\end{equation}
leading to the fluctuation-dissipation theorem:
\begin{equation}
    \hbar\, \chi^{jk}_{\mathcal{O};D}(\omega) = -\frac{1}{2} \left(1 - e^{-\hbar\beta\omega} \right) \mathcal{S}^{jk}_{\mathcal{O}}(\omega).
    \label{app_FD_D}
\end{equation}

Finally, we note that at fixed $\omega$, the dissipative part $\chi^{jk}_{\mathcal{O};D}(\omega)$ satisfies the following symmetry properties:
\begin{eqnarray}
    \Re[\chi^{jk}_{\mathcal{O};D}(\omega)] &=& +\Re[\chi^{kj}_{\mathcal{O};D}(\omega)] = -\Re[\chi^{jk}_{\mathcal{O};D}(-\omega)],\\
    \Im[\chi^{jk}_{\mathcal{O};D}(\omega)] &=& -\Im[\chi^{kj}_{\mathcal{O};D}(\omega)] = +\Im[\chi^{jk}_{\mathcal{O};D}(-\omega)].
    \label{app:chi_sym}
\end{eqnarray}
\subsection{B2: Geometry and fluctuation, dissipation}
\subsubsection{B2.1: SLD operator and geometric tensors}

The Hamiltonian, as well as the thermal density matrix $\rho$, are both functions of the parameter $\bm\phi$. Here, we derive the geometry–fluctuation and geometry–dissipation relations.

The symmetric logarithmic derivative (SLD) operator is introduced to compute the geometric tensors. It is defined implicitly via the Lyapunov equation:
\begin{equation}
    \frac{\partial\rho}{\partial\phi_j} = \mathcal L^j\rho + \rho\mathcal L^j.
\end{equation}

In the eigenbasis, the Hamiltonian and density matrix are diagonal:
\begin{equation}
    K = \sum_m E_m |\Phi_m\rangle\langle\Phi_m|,\quad \rho = \sum_m p_m |\Phi_m\rangle\langle\Phi_m|,\quad \text{with } p_m = \frac{e^{-\beta E_m}}{Z}.
\end{equation}

In this basis, the matrix elements of the SLD operator and the current operator $\mathcal O^j \equiv \partial H/\partial\phi_j$ are given by:
\begin{equation}
    \mathcal L_{mn}^j = \delta_{mn}\frac{1}{2}\frac{\partial\ln p_m}{\partial\phi_j} + i\frac{p_n - p_m}{p_n + p_m} A^j_{mn}, \quad \mathcal O^j_{mn} = \delta_{mn}\frac{\partial E_m}{\partial\phi_j} + i(E_n - E_m)A^j_{mn},
\end{equation}
where $A^j_{mn} = -i\langle \Phi_m| \partial/\partial \phi_j |\Phi_n \rangle$ are the non-Abelian Berry connection coefficients.

We decompose the SLD and current operators into classical and quantum contributions: $\mathcal L^j = \mathcal L^j_C + \mathcal L^j_Q$, and $\mathcal O^j = \mathcal O^j_C + \mathcal O^j_Q$. They are:
\begin{align}
    \mathcal L^j_Q &= \sum_{mn} i\, \frac{p_n - p_m}{p_n + p_m} A^j_{mn} |\Phi_m\rangle\langle\Phi_n|, &
    \mathcal L^j_C &= \sum_m \frac{1}{2}\frac{\partial\log p_m}{\partial\phi_j} |\Phi_m\rangle\langle\Phi_m|, \\
    \mathcal O^j_Q &= \sum_{mn} i(E_n - E_m) A^j_{mn} |\Phi_m\rangle\langle\Phi_n|, &
    \mathcal O^j_C &= \sum_m \frac{\partial E_m}{\partial\phi_j} |\Phi_m\rangle\langle\Phi_m|.
\end{align}

The quantum geometric tensor is defined as the two-point function of the quantum part of the SLD operators:
\begin{equation}
    \mathcal S^{jk} = \langle \mathcal L^j_Q \mathcal L^k_Q \rangle = \mathcal F^{jk} + i\mathcal U^{jk} = \sum_{mn} p_m \tanh^2\left[\frac{\beta}{2}(E_m - E_n)\right] A^j_{mn} A^k_{nm}.
\end{equation}

Using the spectral representation of the two-point function from Eq.~(\ref{app:SpectrumRepS}), we find the relation between fluctuations of the quantum parts of the SLD and those of the current operators, valid for all $\omega$:
\begin{equation}
    \mathcal S^{jk}_{\mathcal O_Q}(\omega) = (\hbar\omega)^2 \coth^2\left(\frac{\hbar\beta\omega}{2}\right) \mathcal S^{jk}_{\mathcal L_Q}(\omega).\label{app:geometry-fluctuation1}
\end{equation}

Importantly, the two-point function of the current operator separates into classical and quantum parts without cross terms $\langle\mathcal O_Q\mathcal O_C\rangle$. Moreover, the classical part yields a delta-function peak at zero frequency:
\begin{equation}
    \mathcal S^{jk}_{\mathcal O}(\omega) = \mathcal S^{jk}_{\mathcal O_Q}(\omega) + \mathcal S^{jk}_{\mathcal O_C}(\omega), \quad \mathcal S^{jk}_{\mathcal O_C}(\omega) \sim \delta(\omega).
\end{equation}
This classical contribution does not enter the dissipation $\chi^{jk}_{\mathcal O; D}(\omega)$ due to the prefactor $1 - e^{-\hbar\beta\omega}$ in the fluctuation–dissipation theorem:
\begin{equation}
    \hbar\chi^{jk}_{\mathcal O; D}(\omega) = -\frac{1}{2} \left(1 - e^{-\hbar\beta\omega}\right) \mathcal S^{jk}_{\mathcal O_Q}(\omega).\label{app:FD}
\end{equation}

We now define the time-dependent geometric tensor as the two-point function $\langle\mathcal L^j_Q(t)\mathcal L^k_Q(0)\rangle$, written in frequency space as:
\begin{equation}
    \mathcal S^{jk}(\omega) \equiv \mathcal S^{jk}_{\mathcal L_Q}(\omega) = \mathcal F^{jk}(\omega) + i\mathcal U^{jk}(\omega).
\end{equation}
Since $ \mathcal S^{jk}(\omega)$ is Hermitian, its real and imaginary parts are, respectively, symmetric and antisymmetric under $j \leftrightarrow k$. Their time-domain counterparts are:
\begin{equation}
    \mathcal F^{jk}(t) = \frac{\mathcal S^{jk}(t) + \mathcal S^{kj}(t)}{2},\quad \mathcal U^{jk}(t) = \frac{\mathcal S^{jk}(t) - \mathcal S^{kj}(t)}{2}.
\end{equation}

Finally, we define the \emph{symmetric} and \emph{antisymmetric} geometric tensors in time domain:
\begin{equation}
    \mathcal S_S^{jk}(t) = \frac{\mathcal S^{jk}(t) + \mathcal S^{kj}(-t)}{2} = \frac{1}{2} \langle\{\mathcal L^j(t), \mathcal L^k(0)\}\rangle,\quad \mathcal S_A^{jk}(t) = \frac{\mathcal S^{jk}(t) - \mathcal S^{kj}(-t)}{2} = \frac{1}{2} \langle[\mathcal L^j(t), \mathcal L^k(0)]\rangle. \label{app_def_SsSa}
\end{equation}
These are the Fourier transforms of $\mathcal S^{jk}_S(\omega)$ and $\mathcal S^{jk}_A(\omega)$ defined in Eq.~(\ref{app_def_Ss_Sa}). Here, $\mathcal S_S^{jk}$ is real and $\mathcal S_A^{jk}$ is imaginary.
\subsubsection{B2.2: Generating function and zeroth order sum rule}
Collecting the results derived above—especially the fluctuation-dissipation theorem Eq.~\eqref{app:FD} and the geometry–fluctuation relation Eq.~\eqref{app:geometry-fluctuation1}—we arrive at the geometry–dissipation relation, which is the key result of this section:
\begin{equation}
    \frac{-1}{2\hbar} \mathcal S^{jk}(\omega) = \frac{\tanh^2\left(\frac{\hbar\beta\omega}{2}\right)}{1-e^{-\hbar\beta\omega}} \frac{\chi^{jk}_{\mathcal O; D}(\omega)}{(\hbar\omega)^2}. \label{app:geometry-dissipation}
\end{equation}

Integrating both sides of the generating function Eq.~\eqref{app:geometry-dissipation} and using the symmetry properties of the dissipative part of the susceptibility—namely, $\Re[\chi_{\mathcal O; D}^{jk}(\omega)] = -\Re[\chi_{\mathcal O; D}^{jk}(-\omega)]$ and $\Im[\chi_{\mathcal O; D}^{jk}(\omega)] = \Im[\chi_{\mathcal O; D}^{jk}(-\omega)]$—we arrive at the zeroth-order sum rules:
\begin{equation}
    -\hbar\pi\mathcal F^{jk} = \int_{0}^{\infty} d\omega ~ \tanh\left(\frac{\hbar\beta\omega}{2}\right) \frac{\Re\chi^{jk}_{\mathcal O; D}(\omega)}{\omega^{2}}, \quad 
    -\hbar\pi\mathcal U^{jk} = \int_{0}^{\infty} d\omega ~ \tanh^2\left(\frac{\hbar\beta\omega}{2}\right) \frac{\Im\chi^{jk}_{\mathcal O; D}(\omega)}{\omega^{2}}.
\end{equation}
Here, $\mathcal F^{jk} = \mathcal F^{jk}(t = 0) = \Re[ \mathcal S^{jk}]$ and $\mathcal U^{jk} = \mathcal U^{jk}(t = 0) = \Im [\mathcal S^{jk}]$ denote the static tensors.

\subsubsection{B2.3: Higher order sum rules}
We multiply the real and imaginary part of Eq.~\eqref{app:geometry-dissipation} by $\omega^n$ and integrate over frequency. This gives,
\begin{eqnarray}
    -\frac{\pi}{\hbar} \mathcal F^{jk}_{(n)} = -\frac{\pi}{\hbar} \int_{-\infty}^{\infty} \frac{d\omega}{2\pi} ~ \omega^n \mathcal F^{jk}(\omega) &=& \int_{-\infty}^{\infty} d\omega ~ \frac{\tanh^2\left(\frac{\hbar\beta\omega}{2}\right)}{1-e^{-\hbar\beta\omega}} \frac{\Re\chi^{jk}_{\mathcal O; D}(\omega)}{\hbar^2\omega^{2-n}},\\
    -\frac{\pi}{\hbar} \mathcal U^{jk}_{(n)} = -\frac{\pi}{\hbar} \int_{-\infty}^{\infty} \frac{d\omega}{2\pi} ~ \omega^n \mathcal U^{jk}(\omega) &=& \int_{-\infty}^{\infty} d\omega ~ \frac{\tanh^2\left(\frac{\hbar\beta\omega}{2}\right)}{1-e^{-\hbar\beta\omega}} \frac{\Im\chi^{jk}_{\mathcal O; D}(\omega)}{\hbar^2\omega^{2-n}},
\end{eqnarray}
where we have introduced,
\begin{equation}
    \mathcal F^{jk}_{(n)} \equiv (i\partial_t)^n\mathcal F^{jk}(t)|_{t=0},\quad \mathcal U^{jk}_{(n)} \equiv (i\partial_t)^n\mathcal U^{jk}(t)|_{t=0}.
\end{equation}

Using the symmetries of the dissipative response function, {\it i.e.} the real/imaginary part is an odd/even function of frequency, we arrive at the $n$th order sum rule: when $n = 2p$,
\begin{equation}
    -\hbar\pi\mathcal F^{jk}_{(2p)} = \int_{0}^{\infty} d\omega ~ \tanh\left(\frac{\hbar\beta\omega}{2}\right) \frac{\Re[\chi^{jk}_{\mathcal O; D}(\omega)]}{\omega^{2-2p}},\quad - \hbar\pi \mathcal U^{jk}_{(2p)} = \int_{0}^{\infty} d\omega ~ \tanh^2\left(\frac{\hbar\beta\omega}{2}\right) \frac{\Im[\chi^{jk}_{\mathcal O; D}(\omega)]}{\omega^{2-2p}},
\end{equation}
and when $n = 2p + 1$,
\begin{equation}
    -\hbar\pi\mathcal F^{jk}_{(2p+1)} = \int_{0}^{\infty} d\omega ~ \tanh^2\left(\frac{\hbar\beta\omega}{2}\right) \frac{\Re[\chi^{jk}_{\mathcal O; D}(\omega)]}{\omega^{2-2p-1}},\quad - \hbar\pi \mathcal U^{jk}_{(2p+1)} = \int_{0}^{\infty} d\omega ~ \tanh\left(\frac{\hbar\beta\omega}{2}\right) \frac{\Im[\chi^{jk}_{\mathcal O; D}(\omega)]}{\omega^{2-2p-1}}.
\end{equation}
\subsection{B3: Case study --- optical conductivity}
\label{subsec: B3: Case study --- optical conductivity}
\subsubsection{B3.1: Electric conductivity tensor and optical absorption}
In the general case, conductivity $\sigma^{ab}(\bm r, t)$ is a real-valued tensor relating a time-dependent and spatially nonuniform electric field to the current,
\begin{equation}
    J^a(\bm r, t) = \int d^d\bm r' \int_{-\infty}^{\infty} dt' ~ \sigma^{ab}(\bm r-\bm r', t-t') E_b(\bm r', t').
\end{equation}

To proceed, we take the Coulomb gauge. In this gauge (we set the speed of light $c=1$), the electric field and the vector potential are, in the frequency domain, related as follows
\begin{equation}
    E_a(\bm r, \omega) = i\omega A_a(\bm r, \omega).
\end{equation}

The perturbation enters the system via the minimal coupling,
\begin{equation}
    \delta H = -\int d^d\bm r ~ \mathcal{J}^a(\bm r) A_a(\bm r, t) + \dots, \label{app_JA}
\end{equation}
where the higher-order terms responsible for diamagnetic current are contained in ``$\dots$''. The Kubo formula dictates
\begin{equation}
    J^a(\bm r, t) = \int d^d\bm r' \int_{-\infty}^{\infty} dt' ~ \chi_{\mathcal J}^{ab}(\bm r-\bm r', t-t') A_b(\bm r', t') + K^{ab} A_b(\bm r, t),
\end{equation}
where $\chi_{\mathcal J}^{ab}(t)$ is the retarded current-current correlation function, and $K^{ab}$ is the diamagnetic response, which is real-valued, static, and symmetric. With the above, the conductivity tensor becomes
\begin{equation}
    \sigma^{ab}(\bm q, \omega) = \frac{i}{\omega}\left[\chi_{\mathcal J}^{ab}(\bm q, \omega) + K^{ab}\right], \quad \chi_{\mathcal J}^{ab}(\bm r-\bm r', t-t') = -\frac{i}{\hbar}\Theta(t - t')\langle\left[\mathcal J^a(\bm r, t), \mathcal J^b(\bm r', t')\right]\rangle.
\end{equation}

Optical conductivity is the response of current to a spatially uniform electric field. Thus, one can set $\bm q = 0$ in the above discussion. The dissipative part of optical conductivity represents the absorption rate of light by the material. The concrete form of the dissipative response differs slightly from the general discussion in Eq.~\eqref{app_general_dissipation}, because in optical conductivity the perturbation source is the vector potential $A_a$ and the actual gauge-invariant probe is the electric field $E_a$.

The energy absorption rate is
\begin{equation}
    \frac{d}{dt}W(t) = -J^a(t)\dot A_a(t) = J^a(t)E_a(t),
\end{equation}
and the total dissipation is
\begin{equation}
    \int_{-\infty}^{\infty} \frac{d}{dt}W(t)\,dt = \int_{-\infty}^{\infty} dt \int_{-\infty}^{\infty} dt' \sigma^{ab}(t-t')E_a(t)E_b(t') = \int_{-\infty}^{\infty} dt \int_{-\infty}^{\infty} dt' \sigma^{ab}_D(t-t')E_a(t)E_b(t'),
\end{equation}
where $\sigma^{ab}_D(t) \equiv [\sigma^{ab}(t) + \sigma^{ba}(-t)]/2$ is the dissipative (absorptive) part of optical conductivity. It is crucial to note that the diamagnetic response does not contribute to optical absorption; the absorption is determined solely by the current-current response:
\begin{equation}
    \sigma^{ab}_D(\omega) \equiv \lim_{\bm q \rightarrow 0} \sigma^{ab}_D(\bm q, \omega) = \lim_{\bm q \rightarrow 0} \frac{i}{2\omega}\left[\chi^{ab}_{\mathcal J}(\bm q, \omega) - \chi^{ba}_{\mathcal J}(\bm q, -\omega)\right] = \lim_{\bm q \rightarrow 0} \frac{-1}{\omega}\chi^{ab}_{\mathcal J; D}(\bm q, \omega).
\end{equation}

In the following, we will omit the $\bm q \rightarrow 0$ label for simplicity in discussing optical conductivity. We will also relate it to the geometry in twisted boundary condition space.
\subsubsection{B3.2: Preliminary for deriving geometric sum rules --- current operators and relations}

To fit into the framework of the general discussion presented in Section~B2, the parameter space for optical conductivity will be the twisted boundary condition space (also called flux space, or equivalently quasi-momentum/many-body momentum space). Consider first a single Bloch wavefunction whose Bloch momentum is quantized on an $N_1 \times N_2 \times \dots \times N_d$ lattice, shifted by the twist angles $\alpha_{1},\dots, \alpha_{d}$, on a given finite size system,
\begin{equation}
    \psi_{\bm k}(\bm r+\bm a_j) = e^{i\bm k\cdot\bm a_j}\psi_{\bm k}(\bm r),\quad \bm k = \sum_{j=1}^{d} \left(m_j + \frac{\alpha_j}{2\pi}\right) \frac{\bm b^j}{N_j}.
\end{equation}

Here, $\bm L_j = N_j\bm a_j$ spans the entire system in the $j$-th direction. The $\bm a_j$ and $\bm b^j$ are the lattice and reciprocal lattice vectors, respectively. The twisted boundary condition $\alpha_j$ determines the single-particle Hilbert space and imposes the boundary condition for the many-body state $\Psi_m$: translating any particle across the system yields
\begin{eqnarray}
    \Psi_m(\bm r_1,\dots,\bm r_n+\bm L_j,\dots,\bm r_{N_e}) &=& e^{i\alpha_j}\Psi_m(\bm r_1,\dots,\bm r_n,\dots,\bm r_{N_e}) \nonumber \\
    &=& e^{i\bm K \cdot \bm L_j}\Psi_m(\bm r_1,\dots,\bm r_n,\dots,\bm r_{N_e}),
\end{eqnarray}
where the many-body momentum is given by
\begin{equation}
    \bm K = \sum_{j=1}^{d} \frac{\alpha_j}{2\pi}\frac{\bm b^j}{N_j}. \label{app_manybodyK}
\end{equation}

There are various names for $\alpha_j$ and $\bm K$: $\alpha_j$ is also referred to as a flux, while $\bm K$ can be called the quasi-momentum. We will use “twisted boundary condition space” and “flux space” interchangeably.

Through the minimal coupling $\bm k \rightarrow \bm p - e\bm A(t)/\hbar$, the flux $\alpha_j$ directly determines a vector potential:
\begin{equation}
    \bm A(t)\cdot\bm L_j = \sum_{a=1}^{d} A_a(t) L^a_j \equiv -\alpha_j(t)\frac{\hbar}{e},\quad \bm A(t) = -\frac{\hbar}{e}\sum_{j=1}^{d} \frac{\alpha_j(t)}{2\pi}\frac{\bm b^j}{N_j}. \label{app_flux}
\end{equation}

Here $e = -|e|$ is the electric charge. Thus, the electric field tunes the twisted boundary condition, explaining how optical conductivity relates to flux space geometry. The perturbation term Eq.~\eqref{app_JA} (to leading order in $A_a(t)$) becomes
\begin{equation}
    \delta H = -\mathcal J^a_T A_a(t) = \sum_{j=1}^{d} \mathcal O^j \alpha_j + \dots, \label{app_deltaH_JO}
\end{equation}
where the total current operator $\mathcal J^a_T \equiv \int d^d\bm r \mathcal J^a(\bm r)$ is introduced.

Regarding $ b_a^j/N_j$ as a matrix with left index $a$ and right index $j$, its inverse can be defined. The left-inverse is $ L^a_j/2\pi$, and the right-inverse is denoted by $\Lambda^a_j$:
\begin{equation}
    \sum_{a=1}^{d} \left(\frac{L^a_j}{2\pi}\right) \left(\frac{b_a^k}{N_k}\right) = \delta^k_j,\quad \sum_{j=1}^{d} \left(\frac{b_a^j}{N_j}\right) \Lambda^b_j = \delta^a_b. \label{app_dualbasis}
\end{equation}

From Eqs.~\eqref{app_flux}, \eqref{app_deltaH_JO}, and \eqref{app_dualbasis}, we obtain the relation between the generalized current $\mathcal O^j$ and the total current operator $\mathcal J^a_T$:
\begin{equation}
    \mathcal O^j = \frac{\hbar}{2\pi e}\sum_{a=1}^{d}\mathcal J^a_T \frac{b_a^j}{N_j},\quad \mathcal J^a_T = \frac{2\pi e}{\hbar}\sum_{j=1}^{d} \Lambda^a_j \mathcal O^j. \label{app_optical_Oi}
\end{equation}

Likewise, the geometry in flux space $\mathcal S^{jk}$ and in many-body momentum space $\mathcal S^{ab}$ are related by:
\begin{equation}
    \frac{\partial}{\partial K_a} = 2\pi \sum_{j=1}^{d} \Lambda^a_j \frac{\partial}{\partial\alpha_j},\quad \frac{\partial}{\partial\alpha_j} = \frac{1}{2\pi}\sum_{a=1}^{d} \frac{b_a^j}{N_j} \frac{\partial}{\partial K_a},
\end{equation}
yielding
\begin{equation}
    \mathcal S^{ab}(\omega) = (2\pi)^2 \sum_{j,k=1}^{d} \Lambda^a_j \Lambda^b_k \mathcal S^{jk}(\omega),\quad \mathcal S^{jk}(\omega) = \frac{1}{(2\pi)^2} \sum_{a,b=1}^{d} \frac{b_a^j}{N_j} \frac{b_b^k}{N_k} \mathcal S^{ab}(\omega). \label{app_Sjk_Sab}
\end{equation}

Finally, the total current operator is also related to the total coordinate operator. The local current is expressed via the velocity operator $\mathcal V_n = \frac{i}{\hbar}[H, \mathcal R_n]$ as:
\begin{equation}
    \mathcal J^a(\bm r) = \frac{e}{2} \sum_{n=1}^{N_e} \left[\mathcal{V}^a_n \delta(\bm r-\bm r_n) + \delta(\bm r-\bm r_n)\mathcal{V}^a_n\right].
\end{equation}
Hence, the total current operator is
\begin{equation}
    \mathcal J_T^a = \int d^d\bm r ~ \mathcal J^a(\bm r) = \frac{ie}{\hbar} [H, \mathcal R^a] = \frac{e}{\hbar}\frac{\partial H}{\partial K_a}. \label{app_JT}
\end{equation}

Its matrix elements are
\begin{equation}
    \left(\mathcal R^a\right)_{mn} = -A^a_{mn},\quad \left(\mathcal J_T^a\right)_{mn} = \frac{ie}{\hbar}(E_n - E_m) A^a_{mn}, \label{app_matrixRJ}
\end{equation}
where $A^a_{mn} = -i\langle\Phi_m|\partial/\partial K_a |\Phi_n\rangle$ is the non-Abelian Berry connection in many-body momentum space. These follow from $[\mathcal R^a, K_b] = i\delta^a_b$, which implies $\mathcal R^a = i\partial/\partial K_a  = 2\pi i \sum_j \Lambda^a_j\partial/\partial \alpha_j$.
\subsubsection{B3.3: Geometrical sum rules related to optical conductivity}
Following the general results discussed in Section.~B.2, and setting the twisted boundary condition $\alpha_j$ at the parameter of consideration, one arrives at the generating function relating flux space geometry $\mathcal S^{jk}(\omega)$ and the dissipative susceptibility $\chi^{jk}_{\mathcal O; D}$,
\begin{equation}
    -\frac{1}{2\hbar} \mathcal S^{jk}(\omega) = \frac{\tanh^2\left(\frac{\beta\omega}{2}\right)}{1-e^{-\hbar\beta\omega}} \frac{\chi^{jk}_{\mathcal O; D}(\omega)}{\hbar^2\omega^2}. \label{app:geometry-dissipation_copy}
\end{equation}

The inverse function Eqn.~(\ref{app_dualbasis}) allows one to replace the flux space geometry to that of many-body space,
\begin{eqnarray}
    \sum_{jk} \Lambda_j^a \Lambda_k^b \chi^{jk}_{\mathcal O; D}(\omega) &=& \left(\frac{\hbar}{2\pi e}\right)^2 \chi^{ab}_{\mathcal J_T; D}(\omega),\nonumber\\
    &=& V \left(\frac{\hbar}{2\pi e}\right)^2 \int d^d\bm r ~ \chi^{ab}_{\mathcal J; D}(\bm r, \omega) = V \left(\frac{\hbar}{2\pi e}\right)^2 \chi^{ab}_{\mathcal J; D}(\bm q \rightarrow \bm 0, \omega),\nonumber\\
    &=& -\omega V \left(\frac{\hbar}{2\pi e}\right)^2 \sigma^{ab}_{D}(\omega),
\end{eqnarray}
where in deriving the above, the following step is used,
\begin{eqnarray}
    \chi^{ab}_{\mathcal J_T}(t-t') &=& \frac{-i}{\hbar}\Theta(t-t')\langle\left[\mathcal J_T(t), \mathcal J_T(t')\right]\rangle,\nonumber\\
    &=& \int_V d^d\bm r \int_V d^d\bm r' ~ \frac{-i}{\hbar}\Theta(t-t')\langle\left[\mathcal J(\bm r, t), \mathcal J(\bm r', t')\right]\rangle,\nonumber\\
    &=& V \int_V d^d\bm r ~ \chi^{ab}_{\mathcal J}(\bm r, t-t').
\end{eqnarray}

Plugging this into the generating function Eqn.~(\ref{app:geometry-dissipation_copy}) gives the generating function relating the optical absorption and twisted boundary condition space geometry,
\begin{equation}
    \frac{2\pi^2}{V}\frac{e^2}{\hbar}S^{jk}(\omega) = \frac{\tanh^2(\hbar\beta\omega/2)}{1-e^{-\hbar\beta\omega}} \sum_{a,b=1}^{d} \frac{\sigma^{ab}_D(\omega)}{\omega} \frac{b_a^j}{N_j} \frac{b_b^k}{N_k},\label{app_genfun_general}
\end{equation}
or, equivalently, relating to the many-body momentum space geometry,
\begin{eqnarray}
    \frac{1}{2V} \frac{e^2}{\hbar} \mathcal S^{ab}(\omega) = \frac{\tanh^2(\hbar\beta\omega/2)}{1-e^{-\hbar\beta\omega}} \frac{\sigma^{ab}_D(\omega)}{\omega}.\label{app_genfun_general_K}
\end{eqnarray}

The zeroth order sum rules are obtained by integrating frequency in Eqn.~(\ref{app_genfun_general_K}),
\begin{equation}
    \frac{\pi}{V} \frac{e^2}{\hbar} \mathcal{F}^{ab} = \int_{0}^{\infty} d\omega ~ \tanh\left(\frac{\hbar\beta\omega}{2}\right) \frac{\Re\sigma^{ab}_{D}(\omega)}{\omega},\quad \frac{\pi}{V} \frac{e^2}{\hbar} \mathcal{U}^{ab} = \int_{0}^{\infty} d\omega ~ \tanh^2\left(\frac{\hbar\beta\omega}{2}\right) \frac{\Im\sigma^{ab}_{D}(\omega)}{\omega},\label{app_genfun_general}
\end{equation}
where they were initially derived in Refs.~\cite{Hauke2016,Carollo_SciRep19} by comparing the respective spectral representations.

The above geometrical sum rules lead to various important consequences, including quantization of Hall conductivity, localization length, orbital magnetization sum rules. In below, we briefly review the sum rules for Hall conductivity and orbital magnetization.

\emph{Zero temperature static 2D Hall conductivity. --- }
The zero temperature limit of the sum rule for the curvature part relates static Hall conductivity to flux space curvature, and for gapped system in two dimension it implies the quantization of $\sigma_H$. The imaginary part sum rule in two dimension at zero temperature is,
\begin{eqnarray}
    \frac{\pi}{V}\frac{e^2}{\hbar}\Omega^{ab} &=& \int_{-\infty}^{\infty} d\omega ~ \frac{\Im\sigma^{ab}_{D}(\omega, T = 0)}{\omega},\nonumber\\
    &=& \int_{-\infty}^{\infty} d\omega ~ \frac{\Im\sigma^{ab}(\omega, T = 0) + \Im\sigma^{ba}(-\omega, T = 0)}{2\omega} = \int_{-\infty}^{\infty} d\omega ~ \frac{\Im\sigma^{ab}(\omega, T = 0) - \Im\sigma^{ba}(\omega, T = 0)}{2\omega},
\end{eqnarray}
where the properties that $\Im\sigma_D^{ab}(\omega)$ is anti-symmetric and odd with to $\omega \leftrightarrow -\omega$ are used. Because $\Im\sigma_D^{ab}(\omega)$ is odd, it vanishes identically at zero frequency. According to the Cauchy's residue theorem,  the contribution of the singular point $\omega=0$ in the integrand is zero. Thus, one can replace the last line with its principal value part,
\begin{equation}
    \frac{\pi}{V}\frac{e^2}{\hbar}\Omega^{ab} = \mathcal P \int_{-\infty}^{\infty} d\omega ~ \frac{\Im\sigma^{ab}(\omega, T = 0) - \Im\sigma^{ba}(\omega, T = 0)}{2\omega}.
\end{equation}

The Kramers-Kronig relation states that the real and imaginary parts of the zero-temperature optical conductivity $\sigma^{ab}(\omega)$ are related to each other via
\begin{equation}
    \Re \sigma^{ab}(\omega) = \mathcal P \int_{-\infty}^{\infty} \frac{d\omega'}{\pi} \frac{\Im \sigma^{ab}(\omega')}{\omega' - \omega}, \quad \Im \sigma^{ab}(\omega) = -\mathcal P\int_{-\infty}^{\infty} \frac{d\omega'}{\pi} \frac{\Re \sigma^{ab}(\omega')}{\omega' - \omega}.
\end{equation}
By applying the above Kramers-Kronig relation, we have,
\begin{equation}
    \frac{\pi}{V}\frac{e^2}{\hbar}\Omega^{ab} = \frac{\pi}{2} \left[\Re\sigma^{ab}(\omega = 0, T = 0) - \Re\sigma^{ba}(\omega = 0, T = 0)\right] = \pi \epsilon^{ab} \sigma_H,
\end{equation}
where $\sigma_H$ is the static Hall conductivity. Therefore, we arrive at the flux space Berry curvature,
\begin{equation}
    \frac{4\pi^3}{V}\frac{e^2}{\hbar}\Omega^{jk} = \pi \sigma_H \sum_{a,b=1}^{d=2} \epsilon^{ab} \frac{b_a^j}{N_j} \frac{b_b^k}{N_k} = \pi \sigma_H \sum_{a,b=1}^{d=2} \frac{\bm b^j \times \bm b^k}{N_j N_k} = \frac{\epsilon^{jk} 4\pi^3}{V} \sigma_H,
\end{equation}
where in deriving the above $V = 2\pi N_1 N_2 S$ is used. We have taken the area of unit cell to be $|\bm a_1\times\bm a_2| = 2\pi S$ hence accordingly the area of reciprocal lattice unit cell is $|\bm b^1\times\bm b^2| = 2\pi/S$. The above result gives Hall conductivity as twisted boundary condition space curvature,
\begin{equation}
    \sigma_H = \frac{e^2}{\hbar} \Omega^{12}.
\end{equation}
When the system is gapped, integrating $\Omega^{jk}$ over the twisted boundary space gives the many-body Chern number $\mathcal C$ which is quantized to integer,
\begin{equation}
    \frac{1}{2\pi}\int_0^{2\pi}d\alpha_1\int_0^{2\pi}d\alpha_2 ~ \Omega^{12}(\alpha_1, \alpha_2) = \mathcal C.
\end{equation}
Since in the thermodynamic limit, $\Omega^{jk}(\alpha_1, \alpha_2)$ become independent on the boundary condition, we arrives at $\Omega^{12} = \mathcal C/2\pi$. As a consequence, the zero temperature static Hall conductivity is quantized in gapped systems,
\begin{equation}
    \sigma_H = \frac{e^2}{\hbar} \Omega^{12} = \frac{e^2}{h} \mathcal C.
\end{equation}
\subsubsection{B3.4: Orbital magnetization sum rules}

We introduce the ``time-dependent orbital magnetization'', valid in dimension $d=2$, as
\begin{equation}
    \mathcal M(t) \equiv \varepsilon_{ab}\langle\mathcal R^a(t)\mathcal J^b(0)\rangle = \varepsilon_{ab}\langle\mathcal R^a(t)\mathcal J_T^b(0)\rangle/V,
\end{equation}
which reduces to the static orbital magnetic moment in the static limit. Using Eq.~\eqref{app_matrixRJ}, the spectral form of $\mathcal M(t)$ becomes
\begin{eqnarray}
    V\mathcal M(\omega) &=& 2\pi \varepsilon_{ab} \sum_{mn} p_m \mathcal R^a_{mn} \mathcal J^b_{T; nm} \delta\left(\omega - E_{nm}/\hbar\right),\nonumber\\
    &=& 2\pi i e \varepsilon_{ab} \sum_{mn} p_m A^a_{mn}A^b_{nm}\omega \delta\left(\omega - E_{nm}/\hbar\right).
\end{eqnarray}
On the other hand, we have
\begin{eqnarray}
    i\omega \varepsilon_{ab} \mathcal S^{ab}(\omega) &=& 2\pi i \epsilon_{ab} \sum_{mn} \omega \; p_m \left(\frac{p_m - p_n}{p_m + p_n}\right)^2 A^a_{mn} A^b_{nm} \delta\left(\omega - E_{nm}/\hbar\right),\\
    &=& 2\pi i \varepsilon_{ab} \sum_{mn} p_m A^a_{mn} A^b_{nm}  \omega \;\tanh^2\left(\frac{\hbar\beta\omega}{2}\right) \delta\left(\omega - E_{nm}/\hbar\right).
\end{eqnarray}
By comparing $\mathcal M(\omega)$ and $\mathcal S^{ab}(\omega)$ and using Eq.~\eqref{app_genfun_general_K}, we arrive at
\begin{equation}
    \mathcal M(\omega) = (e/V) \varepsilon_{ab}\coth^2\left(\frac{\hbar\beta\omega}{2}\right) [i\omega\mathcal S^{ab}(\omega)] = -\frac{\hbar}{e}\varepsilon_{ab}\frac{2\Im[\sigma^{ab}_D(\omega)]}{1-e^{-\hbar\beta\omega}}, \label{app_orbitalM_generatingfun}
\end{equation}
which, after integrating over frequency, gives
\begin{equation}
    -\frac{\pi e}{\hbar}\mathcal M = \varepsilon_{ab} \int_{0}^{\infty} d\omega ~ \coth\left(\frac{\beta\omega}{2}\right) \Im[\sigma^{ab}_D(\omega)].
\end{equation}

This is the finite-temperature generalization of the magnetic circular dichroic sum rule, originally derived by Kuneš and Oppeneer for systems with open boundary conditions at zero temperature in a many-body framework~\cite{Kunes_2000}, extended to periodic topologically trivial insulators in the single-particle picture by Souza and Vanderbilt~\cite{Souza_Vanderbilt_2008}, and later generalized to arbitrary periodic systems in a many-body formulation by Resta~\cite{Resta_2020}.

In Resta's formulation~\cite{Resta_2020}, the magnetic circular dichroic sum rule at zero temperature is
\begin{eqnarray}
    I^{ab} &=& \int_{0}^{\infty}d\omega\,\Im[\sigma^{ab}_{D}(\omega,T=0)] \\
    &=& \frac{\pi e^2}{\hbar^2V} \Im \langle\partial_{K_a} \Psi_0 \vert (\hat{H}-E_0)\vert\partial_{K_b}\Psi_0 \rangle \\
    &=& \frac{\pi e^2}{\hbar^2V} \Im \sum_{n} E_{n0}\langle\partial_{K_a} \Psi_0 \vert \Psi_n\rangle \langle\Psi_n \vert\partial_{K_b}\Psi_0 \rangle \\
    &=& \frac{\pi e^2}{\hbar^2V} \Im \sum_{n} E_{n0}A_{0n}^{a}A_{n0}^{b},
\end{eqnarray}
where $\Psi_0$ is the many-body ground state. On the other hand, taking the zero-temperature limit of our sum rule yields
\begin{equation}
    -\frac{\pi e}{\hbar}\mathcal M(T=0) = \epsilon_{ab} \int_{0}^{\infty} d\omega ~ \Im[\sigma^{ab}_D(\omega,T=0)] = \frac{\pi e^2}{\hbar V} \epsilon_{ab} \mathcal U_{(1)}^{ab} (T=0) = \epsilon_{ab} I^{ab}.
\end{equation}
The last equality implies
\begin{equation}
    I^{ab} = \frac{\pi e^2}{\hbar V} \mathcal U_{(1)}^{ab}(T=0),
\end{equation}
which clearly reveals the geometric meaning of $I^{ab}$ introduced in Ref.~\cite{Resta_2020}.

Importantly, our sum rules are valid at both zero and finite temperatures. At finite temperature, the sum rule for the geometric tensor reads
\begin{equation}
    \frac{\pi}{V} \frac{e^2}{\hbar} \mathcal U_{(1)}^{ab} = \int_{0}^{\infty} d\omega ~ \tanh\left(\frac{\hbar\beta\omega}{2}\right) \Im[\sigma^{ab}_{D}(\omega)],
\end{equation}
while the orbital magnetization sum rule becomes
\begin{equation}
    -\frac{\pi e}{\hbar}\mathcal M = \varepsilon_{ab} \int_{0}^{\infty} d\omega ~ \coth\left(\frac{\beta\omega}{2}\right) \Im[\sigma^{ab}_D(\omega)].
\end{equation}

Note that $\mathcal M$ as defined here is not the full orbital magnetization, but only its geometrical component involved in the sum rule. The total orbital magnetization of correlated many-body systems with periodic boundary conditions remains an open problem~\cite{Resta_2020}. Below we discuss its relation to the total orbital magnetization $\bm{M}_{\text{T}}$ at the single-particle level.

In the single-particle case, the magnetic circular dichroic sum rule measures the orbital magnetic moment arising from the self-rotation of a wavepacket in the semiclassical picture~\cite{Ming-Che_Zhang_1996,Xiao_2010}. Replacing many-body quantities in $\mathcal M(t=0)$ by their single-particle counterparts, i.e., $p_{n}\rightarrow f_{n\bm{k}}$, $A^a_{mn}\rightarrow A^{a}_{mn}(\bm{k})$, $E_{n}\rightarrow\epsilon_{n\bm{k}}$, we obtain
\begin{eqnarray}
    -\frac{1}{2}V\mathcal M^{\text{1p}} 
    &=& i\frac{e}{2\hbar}  \epsilon_{ab} \sum_{mn \bm k} f_{n\bm{k}} (\epsilon_{n\bm k}-\epsilon_{m\bm k}) A^a_{nm}(\bm k)A^b_{mn}(\bm k) \\
    &=& i\frac{e}{2\hbar} \epsilon_{ab}\sum_{n\bm{k}}f_{n\bm{k}}\langle\partial_{k_a}u_{n\bm{k}}|[\epsilon_{n\bm{k}}-\hat{H}(\bm{k})]|\partial_{k_b}u_{n\bm{k}}\rangle \\
    &=& \sum_{n\bm{k}} f_{n\bm{k}} m_{z,n}(\bm{k}),
\end{eqnarray}
where $\bm{m}_{n}(\bm{k}) \equiv (e/2\hbar)i\langle\bm{\nabla}_{\bm{k}}u_{n\bm{k}}|[\epsilon_{n}(\bm{k})-\hat{H}(\bm{k})]\times|\bm{\nabla}_{\bm{k}}u_{n\bm{k}}\rangle$
is the orbital moment for the Bloch state $\vert u_{n\bm{k}}\rangle$, and $\hat{H}(\bm{k})$ is its Hamiltonian~\cite{Xiao_2010}.

The orbital moment involved in the sum rule accounts for only part of the total orbital magnetization $\bm{M}_{\text{T}}$, which is defined by
\begin{equation}
\bm{M}_{\text{T}} = -\frac{1}{V}\left(\frac{\partial\Omega}{\partial\bm{B}}\right)_{T,\mu},
\end{equation}
where $\Omega=E-TS-\mu N$ is the grand potential and $\bm{B}$ is the magnetic field coupled to orbital motion. For a non-interacting electron system, Shi et al.~\cite{Shi_2007} derived the total orbital magnetization using quantum linear response theory, obtaining
\begin{equation}
\bm{M}_{\text{T}}^{\text{1p}} = \sum_{n\bm{k}} f_{n\bm{k}} \bm{m}_{n}(\bm{k}) + \sum_{n\bm{k}} \frac{e}{\hbar \beta} \bm{\Omega}_{n}(\bm{k}) \ln\left(1+e^{-\beta(\epsilon_{n\bm{k}}-\mu)}\right),
\end{equation}
where $\bm{\Omega}_{n}(\bm{k}) \equiv i\langle\bm{\nabla}_{\bm{k}}u_{n\bm{k}}|\times|\bm{\nabla}_{\bm{k}}u_{n\bm{k}}\rangle$ is the Berry curvature. The additional term, not captured by the sum rule, arises from phase-space density-of-states corrections in the semiclassical picture~\cite{Xiao_2005,Xiao_2006}.
\subsubsection{B3.5: Summary of symmetry properties}
We consider a collection of operators $\mathcal O^a$, $a=1,\dots, d$, and summarize the notations and formulas in below. In Table.~\ref{TableDef}, we summarize the definition of response function, dissipation and geometric quantities.
\begin{table}
\begin{center}
\begin{tabular}{ p{3cm} | p{6cm} | p{3cm} | p{5cm} } 
\multirow{2}{2cm}{Response} & response function & $\chi^{ab}_{\mathcal O}(t)$ & $-i\Theta(t)\langle\left[\mathcal O^a(t), \mathcal O^b(0)\right]\rangle$ \\ 
& dissipation & $\chi^{ab}_{\mathcal O; D}(t)$ & $\left[\chi^{ab}_{\mathcal O}(t) - \chi^{ba}_{\mathcal O}(-t)\right]/2i$ \\ \hline 
\multirow{4}{2cm}{Optical conductivity} & current operator & $\mathcal J^a$ & $i[\mathcal R^a, K]$ \\ 
& optical conductivity & $\sigma^{ab}(\omega)$ & $i\left[\chi^{ab}_{\mathcal J}(\omega) - \chi^{ab}_{\mathcal J}(0)\right]/\omega$ \\ 
& absorption part & $\sigma^{ab}_D(\omega)$ & $-\chi^{ab}_{\mathcal J; D}(\omega)/\omega$ \\
& absorption part & $\sigma^{ab}_D(t)$ & $\left[\sigma^{ab}(t) + \sigma^{ba}(-t)\right]/2$ \\ \hline
\multirow{2}{2cm}{Correlation} & correlation function & $\mathcal S^{ab}_{\mathcal O}(t)$ & $\langle\mathcal O^a(t)\mathcal O^b(0)\rangle$ \\
& symmetric and anti-symmetric part & $\mathcal S^{ab}_{\mathcal O; S/A}(\omega)$ & $\left[\mathcal S^{ab}_{\mathcal O}(\omega) \pm \mathcal S^{ba}_{\mathcal O}(-\omega)\right]/2$ \\ \hline 
\multirow{4}{2cm}{Geometry} & symmetric logarithmic derivative & $\mathcal L^a$ & $d\rho = \mathcal L\rho + \rho\mathcal L$ \\ 
& FT-QGT & $\mathcal S^{ab}$ & $\mathcal S^{ab}(t)$ \\ 
& static FT-QGT & $\mathcal S^{ab}$ & $\mathcal S^{ab}(t=0)$ \\
& QFIM and MUC & $\mathcal F^{ab}, \mathcal U^{ab} \in \mathbb{R}$ & $\mathcal S^{ab} = \mathcal F^{ab} + i\mathcal U^{ab}$ 
\end{tabular}
\caption{Summary of definitions.}\label{TableDef}
\end{center}
\end{table}

In Table.~\ref{TableSym}, we summarize symmetry properties of susceptibilities and correlation functions.
\begin{table}
\begin{center}
\begin{tabular}{ p{5cm} | p{5cm} | p{5cm} }
& $a \leftrightarrow b$ & $\omega \leftrightarrow -\omega$\\ \hline \hline
$\Re[\sigma^{ab}(\omega)]$ & N.A. & $+1$\\
$\Im[\sigma^{ab}(\omega)]$ & N.A. & $-1$\\ \hline
$\Re[\sigma^{ab}_{D}(\omega)]$ & $+1$ & $+1$\\
$\Im[\sigma^{ab}_{D}(\omega)]$ & $-1$ & $-1$ \\ \hline \hline
$\Re[\mathcal S^{ab}_{\mathcal O}(\omega)]$ & $+1$ & N.A.\\
$\Im[\mathcal S^{ab}_{\mathcal O}(\omega)]$ & $-1$ & N.A.\\ \hline
$\Re[\mathcal S^{ab}_{\mathcal O; S}(\omega)]$ & $+1$ & $+1$\\
$\Re[\mathcal S^{ab}_{\mathcal O; A}(\omega)]$ & $+1$ & $-1$\\
$\Im[\mathcal S^{ab}_{\mathcal O; S}(\omega)]$ & $-1$ & $-1$\\
$\Im[\mathcal S^{ab}_{\mathcal O; A}(\omega)]$ & $-1$ & $+1$
\end{tabular}
\caption{Summary of symmetry properties. Here ``N.A.'' means the quantity does not go back to itself up to $\pm1$ under this operation.}\label{TableSym}
\end{center}
\end{table}

\section{C: Bounds and Inequalities for the Time-dependent Geometric Tensor}
\subsection{C1: Cauchy-Schwarz type inequality}
We consider a density matrix $\rho$, and two general operators $A, B$. The operators are not necessarily Hermitian. We first of all observe that, since $\rho$ is positive semidefinite, for general operators $A,B$, the following Cauchy-Schwarz-type inequality holds,
\begin{equation}
\text{Tr}\left[\rho A^{\dagger}B\right] \text{Tr}\left[\rho B^{\dagger} A\right] \le\text{Tr}\left[\rho A^{\dagger}A\right] \text{Tr}\left[\rho B^{\dagger}B\right].\label{eq:Inequality_mixed-1}
\end{equation}
In the operators $A, B$ are further Hermitian, it then follows that
\begin{align}
    \det\left[\begin{array}{cc}
    0 & -i\text{Tr}\left[\rho\left[A,B\right]\right]\\
    -i\text{Tr}\left[\rho\left[B,A\right]\right] & 0
    \end{array}\right] & \le\det\left[\begin{array}{cc}
    \text{Tr}\left[\rho\left\{ A,A\right\} \right] & \text{Tr}\left[\rho\left\{ A,B\right\} \right]\\
    \text{Tr}\left[\rho\left\{ B,A\right\} \right] & \text{Tr}\left[\rho\left\{ B,B\right\} \right]
    \end{array}\right].\label{eq:Inequality_det}
\end{align}

Next, we substitute the operators $A$ and $B$ by $\mathcal L^j(t)$ and $\mathcal L^k(t=0)$, respectively. Following Eq.~\eqref{eq:Inequality_det}, one arrives at the following inequality for time-dependent density matrix geometric tensors,
\begin{align}
    \det\left[\begin{array}{cc}
    0 & -i\text{Tr}\left[\rho\left[\mathcal{L}^{j}(t),\mathcal{L}^{k}(0)\right]\right]\\
    -i\text{Tr}\left[\rho\left[\mathcal{L}^{k}(0),\mathcal{L}^{j}(t)\right]\right] & 0
    \end{array}\right] & \le\det\left[\begin{array}{cc}
    \text{Tr}\left[\rho\left\{ \mathcal{L}^{j}(t),\mathcal{L}^{j}(t)\right\} \right] & \text{Tr}\left[\rho\left\{ \mathcal{L}^{j}(t),\mathcal{L}^{k}(0)\right\} \right]\\
    \text{Tr}\left[\rho\left\{ \mathcal{L}^{k}(0),\mathcal{L}^{j}(t)\right\} \right] & \text{Tr}\left[\rho\left\{ \mathcal{L}^{k}(0),\mathcal{L}^{k}(0)\right\} \right]
    \end{array}\right].
\end{align}
Equivalently, it is,
\begin{align}
    -\det\left[\begin{array}{cc}
    0 & \mathcal{S}_{A}^{jk}(t)\\
    \mathcal{S}_A^{kj}(-t) & 0
    \end{array}\right] & \le\det\left[\begin{array}{cc}
    \mathcal{F}^{jj}(0) & \mathcal{S}_S^{jk}(t)\\
    \mathcal{S}_S^{kj}(-t) & \mathcal{F}^{kk}(0)
    \end{array}\right],
\end{align}
where $\mathcal S_S$ and $\mathcal S_A$ are introduced in Eq.~\eqref{app_def_SsSa}. This gives,
\begin{equation}
|\mathcal S^{jk}(t)|^2 \leq \mathcal S^{jj}(0) \mathcal S^{kk}(0).
\end{equation}

\subsection{C2: Positivity of the Hankel matrix and associated principal minors}
\subsubsection{C2.1: General properties of positive functions}
We start by considering a smooth function \( f : \mathbb{R} \to \mathbb{C} \). For instance, $f(t)$ can be one of the diagonal components of the time dependent geometric tensor $\mathcal S^{ii}(t)$. The function $f(t)$ is said to be ``positive definite function'' if and only if: for any test function $c(t)$ the following integral is non-negative,
\begin{equation}
    \int_{\infty}^{\infty} dt \int_{\infty}^{\infty} dt' ~ c^{*}(t) c(t') f(t - t') \geq 0,\label{Def:Positive_Function}
\end{equation}
and the inequality saturates only when function $c(t) = 0$. It is clear that $\mathcal S^{ii}(t)$ qualify to be a positive function because the thermal density matrix is positive and full rank.

For positive function $f(t)$, it must have a positive measure $\mu(\omega) \equiv f(\omega)d\omega$, where $f(\omega)$ denotes the Fourier transform of $f(t)$. To see this, we Fourier transform the above and arrive at the following,
\begin{equation}
    \int_{-\infty}^{\infty} \vert c(\omega)\vert^2 f(\omega)d\omega = \int_{\infty}^{\infty} \vert c(\omega)\vert^2 d\mu(\omega) \geq 0, \label{PF_Fourier}
\end{equation}
from which we see only when $\mu(\omega) \geq 0$ can ensure the positivity of $f(t)$. This is the Bochner's theorem.

Another property of a positive definite function is that, the associated Hankel matrix and its principal minors are all positive definite. To proceed, we introduce the moment function,
\begin{equation}
    m_n = \int_{-\infty}^{\infty} \omega^n \, d\mu(\omega),
\end{equation}
with which one can construct a Hankel matrix,
\begin{equation}
H_{m,n} = m_{m+n} \quad \Leftrightarrow \quad H = \begin{bmatrix}
m_0 & m_1 & m_2 & \cdots \\
m_1 & m_2 & m_3 & \cdots \\
m_2 & m_3 & m_4 & \cdots \\
\vdots & \vdots & \vdots & \ddots
\end{bmatrix}.
\end{equation}

First of all, it is straightforward to prove the Hankel matrix has to be positive definite. This is because for any complex scalars $c_i$, the following inequality holds,
\begin{equation}
\sum_{m,n\geq 0} c^{*}_m c_n H_{m,n} = \sum_{m,n\geq 0} c^{*}_m c_n m_{m+n} = \sum_{m,n\geq 0} c^{*}_m c_n  \int_{-\infty}^{\infty} \omega^{m+n} \, d\mu(\omega) =   \int_{-\infty}^{\infty} \vert \sum_{n\geq 0} c_n \omega^{n} \vert^2 \, d\mu(\omega) \geq 0.
\end{equation}

Secondly, one can show that any principal minor of the Hankel matrix is also positive definite. The principal minor is the determinant of a square submatrix formed by selecting a subset $I=\{i_0,\ldots,i_{k-1}\} \subset \{0,\ldots,n-1\}$ of rows and the same subset of columns.  The proof is as follows. Let's denote  \( M \in \mathbb{C}^{n \times n} \) an Hermitian and positive semi-definite matrix and \( M_I = M[I,I] \in \mathbb{C}^{k \times k} \) a principal submatrix. $M_I$ can be obtained from $M$ by projection $M_I = P_I^{\dagger} M P_I$, where $P_I$ is an $n\times k$ matrix, which selects coordinates in \( I \). Let \( y \in \mathbb{C}^k \), and define \( x \in \mathbb{C}^n \) such that: $x_j =y_m$ if $j = i_m \in I$ and otherwise $x_j =0$. Then, we have $y^\dagger M_I y = x^\dagger M x \geq 0$. Thus, \( M_I \) is positive semidefinite. Since a positive semi-definite Hermitian matrix has non-negative eigenvalues, the determinant of any principal submatrix is non-negative, i.e., $\det(M_I) \ge0$. In summary, all principal minors of a Hermitian positive semidefinite matrix are non-negative.

Although the above discussion was targeted for scalar functions $f(t)$, the result can be easily generalized to matrix type positives functions $f_{ij}(t)$. Clearly, the time-dependent quantum geometric tensor $\mathcal S^{ij}(t)$ is positive: for any test function $c_i(t)$, the following inequality holds,
\begin{equation}
    \sum_{i,j} \int_{-\infty}^{\infty} dt \int_{-\infty}^{\infty} dt' ~ c^*_i(t) c_j(t') \mathcal S^{ij}(t-t') = {\rm Tr}[\rho X^\dag X] \geq 0,
\end{equation}
where $X = \sum_{i} \int_{-\infty}^{\infty} dt ~ c_i(t)\mathcal L^i(t)$. As a result, the following Hankel matrix, as well as its principal minors, are positive,
\begin{equation}
H_{\mathcal S} = 
\begin{bmatrix}
\begin{bmatrix}
\mathcal S^{11}_{(0)} & \mathcal S^{11}_{(1)}  & \mathcal S^{11}_{(2)}  & \cdots \\
\mathcal S^{11}_{(1)}  & \mathcal S^{11}_{(2)}  & \mathcal S^{11}_{(3)}  & \cdots \\
\mathcal S^{11}_{(2)}  & \mathcal S^{11}_{(3)}  & \mathcal S^{11}_{(4)}  & \cdots \\
\vdots & \vdots & \vdots & \ddots
\end{bmatrix} & \begin{bmatrix}
\mathcal S^{12}_{(0)} & \mathcal S^{12}_{(1)}  & \mathcal S^{12}_{(2)}  & \cdots \\
\mathcal S^{12}_{(1)}  & \mathcal S^{12}_{(2)}  & \mathcal S^{12}_{(3)}  & \cdots \\
\mathcal S^{12}_{(2)}  & \mathcal S^{12}_{(3)}  & \mathcal S^{12}_{(4)}  & \cdots \\
\vdots & \vdots & \vdots & \ddots 
\end{bmatrix} & \cdots\\
\begin{bmatrix}
\mathcal S^{21}_{(0)} & \mathcal S^{21}_{(1)}  & \mathcal S^{21}_{(2)}  & \cdots \\
\mathcal S^{21}_{(1)}  & \mathcal S^{21}_{(2)}  & \mathcal S^{21}_{(3)}  & \cdots \\
\mathcal S^{21}_{(2)}  & \mathcal S^{21}_{(3)}  & \mathcal S^{21}_{(4)}  & \cdots \\
\vdots & \vdots & \vdots & \ddots
\end{bmatrix} & \begin{bmatrix}
\mathcal S^{22}_{(0)} & \mathcal S^{22}_{(1)}  & \mathcal S^{22}_{(2)}  & \cdots \\
\mathcal S^{22}_{(1)}  & \mathcal S^{22}_{(2)}  & \mathcal S^{22}_{(3)}  & \cdots \\
\mathcal S^{22}_{(2)}  & \mathcal S^{22}_{(3)}  & \mathcal S^{22}_{(4)}  & \cdots \\
\vdots & \vdots & \vdots & \ddots 
\end{bmatrix} & \cdots\\
\vdots & \vdots & \ddots 
\end{bmatrix}\label{app_Hankel}
\end{equation}

\subsubsection{C2.2: Consequences for time-dependent geometric tensors}
We now discuss few consequences following the positivity of the above Hankel matrix and its principal minors. We will take sub-matrices of Eqn.~(\ref{app_Hankel}) and state a few consequences following the positivity of the sub-matrices.

\begin{itemize}
\item
First of all, we have the following statement: the even order of the diagonal element of the time-dependent geometric tensor has to be positive,
\begin{equation}
\mathcal S^{ii}_{(2n)} \geq 0,\label{app_positive_Sii}
\end{equation}
valid for any $i$ and $n$. This simply follows from the positivity of the principal minor of order one of the Hankel matrix Eqn.~(\ref{app_Hankel}).

\item
For any matrix constructed from $i,j \in I$ of even order $2n$, the matrix  $\left[S^{ij}_{(2n)}\right]_{i,j\in I}$ is positive definite, and
\begin{equation}
\det_{ij \in I} \mathcal S^{ij}_{(2n)} \geq 0,\quad\forall n.
\label{app_positive_Sij}
\end{equation}

Notably, the $n = 0$ component is just the static quantum geometric tensor for density matrix. Eq.~\eqref{app_positive_Sii} gives the positivity of the diagonal elements of the Fisher information matrix. Eq.~\eqref{app_positive_Sij} reduces to the ``determinant bound''~\cite{Carollo2018},
\begin{equation}
\det \mathcal F \geq \det \mathcal U.
\end{equation}
\item
There is a Cauchy-Schwarz-type inequality. For any $m, n$ and $j, k$, the following holds,
\begin{equation}
    \det\begin{bmatrix}
    \mathcal S^{jj}_{(2m)} & \mathcal S^{jk}_{(m+n)}  \\
    \mathcal S^{kj}_{(m+n)}  & \mathcal S^{kk}_{(2n)} 
    \end{bmatrix} = \mathcal S^{jj}_{(2m)}\mathcal S^{kk}_{(2n)} - \mathcal S^{jk}_{(m+n)}\mathcal S^{kj}_{(m+n)} = \mathcal S^{jj}_{(2m)}\mathcal S^{kk}_{(2n)} - |\mathcal S^{jk}_{(m+n)}|^2 \ge 0.
\end{equation}
The diagonal case $j = k$, the above inequality was covered in Ref.~\cite{Raquel2403} at the zero temperature limit (also extended to allowing $j, k$ to be non-integer). There, only the real part of quantum geometric tensor is considered—that is, their result pertains to $\mathcal{F}$, not to our $\mathcal{S}$.

\end{itemize}

\end{document}